# Absorption spectroscopy of individual single-walled carbon nanotubes

Stéphane Berciaud,[a] Laurent Cognet,[a] Philippe Poulin,[b] R. Bruce Weisman,[c] and Brahim Lounis[a*]

[a]*Centre de Physique Moléculaire Optique et Hertzienne, Université Bordeaux I & CNRS (UMR 5798), 351, cours de la Libération, 33405 Talence Cedex, France*

[b]*Centre de Recherche Paul Pascal – CNRS, Avenue Schweitzer 33600 Pessac, France,*

[c]*Department of Chemistry, Smalley Institute for Nanoscale Science and Technology, and Center for Biological and Environmental Nanotechnology, Rice University, Houston, TX 77005, USA*

*Current methods for producing single-walled carbon nanotubes (SWNTs) lead to heterogeneous samples containing mixtures of metallic and semiconducting species with a variety of lengths and defects. Optical detection at the single nanotube level should thus offer the possibility to examine these heterogeneities provided that both SWNT species are equally well detected. Here, we used photothermal heterodyne detection to record absorption images and spectra of individual SWNTs. Because this photothermal method relies only on light absorption, it readily detects metallic nanotubes as well as the emissive semiconducting species. The first and second optical transitions in individual semicontucting nanotubes have been probed. Comparison between the emission and absorption spectra of the lowest-lying optical transition reveal mainly small Stokes shifts. Side bands in the near-infrared absorption spectra are observed and assigned to exciton-phonon bound states. No such sidebands are detected around the lowest transition of metallic nanotubes.*

Single-walled carbon nanotubes (SWNTs) have attracted much attention for their remarkable physical properties, making them promising for applications in nanotechnologies. The diameter and chiral angle of SWNTs are defined by the two structural indices (n,m), which determine their electronic structure including their metallic or semiconducting nature[1, 2]. The optical properties of SWNTs have received growing interest since the recent observation of near-infrared luminescence from well separated surfactant-suspended semiconducting SWNTs[3]. Subsequent photoluminescence excitation measurements resulted in the precise mapping of the transition energies for a large variety of specific semiconducting structural species[4, 5].

All current methods for producing SWNTs lead to heterogeneous samples containing mixtures of metallic and semiconducting species with a variety of lengths and defects. This diversity of SWNT structures complicates precise spectroscopic characterization. Single particle methods thus appear valuable for eliminating the heterogeneity and inhomogeneity present in bulk SWNT optical spectra[6, 7]. For instance, photoluminescence studies performed on individual SWNTs revealed the single-nanotube linewidth of emission spectra and the presence of spectral variations within a given (*n,m*) type[6, 8, 9]. Luminescence measurements are however limited to semiconducting SWNTs. By contrast, Raman scattering has been used to study individual semiconducting and metallic SWNTs[6, 10, 11], but such experiments remain constrained by the weakness of the signal and the need to use near-resonant laser sources. Rayleigh scattering has also been studied to record optical spectra from individual structure-assigned SWNTs[7, 12]. It allowed probing of the third and fourth optical transitions of semiconducting SWNTs as well as the first and second transitions of metallic ones. So far, this technique has only been applied to studying long, large diameter (~2 nm) SWNTs that are individually suspended over open apertures.

Because the cross-sections for optical absorption decrease less rapidly than those for Rayleigh scattering as the nanotube diameter is reduced[13], a detection method relying on the remarkable absorptive properties of carbon nanotubes would be valuable.

Small diameter SWNTs, which are abundantly produced by the HiPco method[3], are particularly appealing for optical studies because their first optical transitions lie in the visible region for metallic SWNTs and in the near-infrared for semiconducting species.

Here, we report the highly sensitive imaging and absorption spectroscopy of individual small diameter SWNTs, using Photothermal Heterodyne Imaging (PHI)[14, 15]. Because PHI probes light absorption, it enables identification of both semiconducting and metallic SWNTs. The chiral indices of individual semiconducting nanotubes are assigned by analyzing their absorption spectra around their first $S_{11}$ and second $S_{22}$ optical resonances. As expected for semiconducting nanotubes, exciton-phonon sidebands are observed near the first optical resonance. No such sidebands are detected around the lowest $M_{11}$ transition of metallic nanotubes.

Semiconducting and metallic SWNTs can be equivalently regarded as ideal candidates for investigations with absorption based methods such as PHI[14]. First, carbon nanotubes are highly absorptive nano-objects with absorption cross-sections of typically $10^{-18}$ cm$^2$/carbon atom[16] and rapid inter- and intra-band carrier relaxation dynamics[17]. Metallic nanotubes have electron-electron and electron-phonon relaxation times in the subpicosecond range[17, 18], whereas semiconducting species have picosecond-scale nonradiative decay times associated with weak ensemble luminescence quantum yields (ca. $10^{-3}$)[19-21].

Our photothermal imaging and absorption spectroscopy apparatus has been described elsewhere[22]. In brief, it consists of a probe beam (HeNe laser, 633nm, output power of ~ 8 mW) overlaid with a modulated tunable cw absorption beam. For the latter we used either a dye laser (tunable from 2.07 to 2.34 eV) for spectroscopy of metallic SWNTs, or a cw

Ti:sapphire laser (tunable from 1.20 to 1.31 eV or from 1.24 to 1.63 eV) for spectroscopy of semiconducting SWNTs. The absorption and probe beams were both focused onto the sample using a high NA objective. The absorption beam was linearly polarized and had an intensity of $\sim 500\, kW/cm^2$. The dye laser was also used at fixed wavelength (565 nm, i.e. photon energy of ~2.2 eV) and with an intensity of $\sim 20\, kW/cm^2$ to record confocal fluorescence images and emission spectra of individual luminescent SWNTs. In those experiments, the collected luminescence light was split between a silicon avalanche photodiode (APD) and a spectrometer equipped with a cooled Si-CCD camera. Due to the limited spectral sensitivity of Si-based detectors, only semiconducting nanotubes emitting in the range 1.19-1.45 eV were detected.

We used SWNTs grown by the HiPco process and prepared following the procedure introduced by O'Connell *et al.*[3]. Dilute suspensions of SWNTs in aqueous sodium dodecylsulfate (SDS) were spin-coated onto clean microscope cover slips. We chose dilution factors and spinning rates to give a final SWNT density of less than $1\, \mu m^{-2}$. A drop of viscous silicone oil was added on top of the samples to ensure homogeneous heat diffusion.

Figure 1 shows typical luminescence (Fig. 1a and 1b) and corresponding photothermal (Fig. 1c and 1d) images acquired with two orthogonal polarizations of the absorption beam at a fixed wavelength (565 nm). Luminescent SWNTs represent only a small proportion of all observed SWNTs, and all of them co-localize with spots in the photothermal images (black circles on Fig 1). Interestingly, many more spots are detected in the photothermal images with high signal-to-noise ratios (from 10 to >1000), providing a more complete picture of the SWNTs present in the sample. The numerous non-luminescent peaks may arise from metallic SWNTs or may correspond to semiconducting species emitting either with very low quantum yields or at photon energies lower than our detection range. One can also not exclude the possibility that a few spots might arise from SWNTs aggregated into small absorbing bundles.

Considering that only ~20% of all semiconducting SWNTs have peak emission wavelengths within the range of our Si-APD, and that the metallic SWNT abundance is ca. 1/3 of the total, we estimate that photothermal images should exhibit ~10 times more spots than luminescence images, in good agreement with the data presented in Fig. 1. Almost all SWNTs appear as diffraction-limited spots, unresolved by the imaging method (Fig. 1a-d). This is consistent with the length distribution of nanotubes prepared using our ultrasonic dispersion process[19, 20]. However, as shown in Fig. 1e-f micrometer long nanotubes were occasionally observed.

The polarization-dependent absorption of SWNTs is evident in both luminescence and photothermal images (see circled SWNTs in Fig. 1a-d, and Fig. 1e-f); the highest absorption being achieved when the SWNT longitudinal axis is parallel to the incident laser field[6, 7]. The variations in photothermal signal strengths can be explained by three factors: the rather broad distribution of nanotube lengths; their random orientations with respect to the linearly polarized absorption beam; and the varying mismatches between absorption beam wavelength and the transition peak among the numerous SWNT (*n,m*) species. The measured luminescence signal strengths can vary additionally because of the wavelength-dependent response of our Si-APD in the near-infrared and possible differences in numbers of defects and trap states among the nanotubes[10, 19].

Prior to measuring absorption spectra of individual semiconducting SWNTs, we recorded 117 individual nanotube luminescence spectra in the 1.19–1.33eV spectral window (10 s integration time per spectrum; examples are shown in Fig 2a-c). Most luminescence spectra display a Lorentzian shape from which we extract the peak emission energy $S_{11}^E$ corresponding to the radiative recombination of the lowest excitonic state[23, 24]. From the distribution of the $S_{11}^E$ values (data not shown), different sub-populations could be identified with mean values lying at 1.28, 1.25 and 1.20 eV. Based on the spectral assignments of

Bachilo et al.[4] and the dominant species observed in bulk spectrofluorimetry of our sample, we identify these as the (8,3), (6,5) and (7,5) species, respectively. In a second step we recorded photothermal absorption spectra of such nanotubes, by scanning the absorption laser frequency with steps of ~2 meV (100 ms integration time per point). For normalization, the absorption beam power was measured at the sample during the acquisitions.

Figure 2a-c displays examples of luminescence and absorption spectra for three SWNTs with different chiralities, belonging respectively to the (8,3), (6,5), and (7,5) species, respectively. An absorption peak $S_{11}^A$ near the emission peak $S_{11}^E$ is clearly visible for the (8,3) and (6,5) tubes (see insets of Fig. 2a and 2b). The absorption peak of the (7,5) nanotubes can be measured with a different setting of the Ti-Sapphire laser, as exemplified in Fig. 2d. The influence of the cw probe beam was investigated by recording two successive spectra of the same nanotube, using a full and a four-fold attenuated probe intensity (Fig. 2e). No noticeable difference between the two normalized spectra was found, indicating that the probe had a negligible influence on the line shape of the absorption spectra.

The distribution of $S_{11}^A$ values obtained from 62 absorption spectra is presented in Fig. 2f, where the positions of (7,5), (6,5) and (8,3) species are indicated.

We now consider the Stokes shift, $\Delta S_{11} = S_{11}^A - S_{11}^E$, measured on 49 (6,5) and (8,3) nanotubes. We find that this shift lies below our experimental accuracy (~5 meV) for most of the nanotubes (Fig 2g, see also for example insets in Fig 2a-b), in agreement with ensemble measurements[25]. As shown in Fig. 2g, no correlation between the position of the emission peak and the Stokes shift values is observed. Interestingly, ~25% of the SWNTs studied exhibit a larger Stokes shift (from ~10 to ~ 40 meV). Furthermore, absorption spectra of the $S_{11}$ transition are broader than emission lines (by a factor ~2). Those effects can be attributed either to spectral jumps as the absorption and luminescence spectra are not acquired simultaneously, or to the presence of trap states due to structural or chemical defects, or local environmental variations along the nanotube. We expect that such traps might be present initially or might be induced by laser irradiation during measurements[26].

Finally, for all semiconducting species, we observe an absorption side-band lying ~200 meV above the $S_{11}$ transition (Fig. 2a-c). This energy difference matches within experimental error the energy of the Raman-active G band vibration. According to theoretical predictions[27] and room temperature ensemble spectra[28], this sideband is assigned to exciton-phonon bound states. In principle, such a sideband could also be attributed to the second-lowest one-photon-allowed excitonic transition, but its chirality-independent position rules out this possible assignment. This suggests that for linear processes most of the oscillator strength of the $S_{11}$ transition is contained in the lowest optically allowed excitonic state. One should also stress that, although exciton-phonon sidebands are detected on single tubes, exciton-phonon interactions remain in the weak coupling limit[29], as confirmed by the small Stokes shifts measured here. Direct observation of such bound states in the absorption spectra of individual SWNTs supports the excitonic origin of optical resonances in SWNTs[23].

The (6,5), and (8,3) species are difficult to fully discriminate by their first optical transition. However, their second optical transitions, $S_{22}$, are well separated[4]. We thus measured the absorption spectra of such nanotubes near the expected $S_{22}$ position of the (6,5) species (Fig. 3a-b). The observation of an absorption peak in Fig 3a and the absence of any feature in Fig. 3b allow us to unambiguously assign the 3a species as (6,5) and the 3b species as (8,3). Interestingly, and in agreement with bulk measurements[30], we find that $S_{22}$ transitions are systematically broader than $S_{11}$ transitions. This observation is consistent with time resolved photoemission measurements that show faster electronic relaxation with increased excitation photon energy[31].

As mentioned above and illustrated in figure 1, PHI is a remarkable detection method for individual metallic SWNTs. Figure 4a-b presents absorption spectra of two non-luminescent SWNTs with peaks at 2.08 eV and 2.23 eV. Figure 4c shows the distribution of the peak energies found in the energy range of the laser for 23 non-luminescent SWNTs, revealing two subpopulations centred at 2.09 eV and 2.23 eV.

Although it might seem that these peaks could arise from $S_{22}$ transitions of semiconducting SWNTs (e.g. the (8,4) or (9,2) species) having peak emissions $S_{11}^E$ falling outside the spectral range of our detector, the transitions are significantly narrower than all $S_{22}$ absorptions measured here (see Fig. 3a). Moreover, in contrast to the spectra shown in Fig. 2, we observed no exciton-phonon sidebands for any absorption peaks near 2.09 eV. This observation constitutes a strong indication that the spectra shown in Fig. 4 stem from the lowest optical transitions ($M_{11}$) of individual metallic nanotubes, for which excitonic effects are expected to be much weaker than in semiconducting tubes[32]. Using results from ensemble resonant Raman spectra[33, 34] and numerical simulations[35], we assign the lower and higher energy groups to species with $2n + m = 27$ and $2n + m = 24$, respectively.

The use of an appropriate source and detector to monitor absorption and emission in the near-infrared, as well as complementary investigations of the trigonal warping effect[12,35] (a hallmark of metallic SWNTs), will allow unambiguous discrimination of metallic from semiconducting absorbers.

In conclusion, we have demonstrated the efficient detection of individual luminescent and nonluminescent SWNTs using a highly sensitive photothermal detection method. Absorption spectra of the lowest-lying optical transitions of individual SWNTs have been recorded. Our absorption technique has the unique ability to probe any optical absorption transition of individual SWNTs in common environments. We believe that it holds great promise for applications in nanotube characterization and sorting.

**Acknowledgement**

This research was funded by CNRS (ACI Nanoscience and DRAB), Région Aquitaine, Agence Nationale de la Recherche (PNANO program) and the French Ministry for Education and Research (MENRT). RBW acknowledges support from the NSF (CHE-0314270) and the Welch Foundation (C-0807).

**Figure Captions:**

**Figure 1**: (a-b) Luminescence and (c-d) photothermal absorption images of the same region ($12 \times 12 \mu m^2$) of a sample containing micelle-encapsulated SWNTs. (e-f) photothermal absorption images of another region ($8 \times 8 \mu m^2$) of a sample containing micrometer sized nanotubes. The orientation of the absorption beam polarization is indicated by the arrows.

**Figure 2**: Absorption and luminescence spectroscopy of the $S_{11}$ optical transition on individual semiconducting SWNTs. (a-c) Absorption spectra (gray: raw data, and red: smoothed data) and luminescence spectra (gray: raw data, black: Lorentzian fit) of individual (8,3), (6,5), and (7,5) semiconducting nanotubes, respectively. Absorption sidebands at 200 meV above the $S_{11}$ energy can be seen on the three spectra. Insets in (a) and (b): zoom in the absorption spectra around the $S_{11}$ transition and corresponding luminescence spectra. (d)* Absorption spectra (gray line: raw data and red: smoothed data) from another (7,5) nanotube around the $S_{11}$ transition. (e) Absorption spectrum of an individual nanotube recorded with full (black) and attenuated (gray) probe beam power. (f) Distribution of the $S_{11}^A$ values measured from 62 absorption spectra. The positions of the (7,5), (6,5), and (8,3) species are indicated. The black line is a triple Gaussian curve intended to guide the eye. (g) Position of the absorption peak $S_{11}^A$ with respect to the emission peak $S_{11}^E$. The blue dashed line corresponds to a zero Stokes shift. The symbol * in (d), (e) and (f) indicate measurements performed using the absorption laser (Ti-Sapph) optimized for the 1.20-1.31eV tuning range.

**Figure 3:** Absorption spectroscopy of the $S_{22}$ optical transition on individual semiconducting SWNTs. Absorption spectra (gray: raw data, and red: smoothed data) and luminescence spectra (gray: raw data, black: Lorentzian profile fit) for a (6,5) nanotube (a) and a (8,3) nanotube (b). The $S_{22}^A$ peak absorption from the (6,5) nanotube is clearly identified at ~ 2.18

eV, whereas for the (8,3) nanotube, no $S_{22}$ peak is detected in the tuning range of the dye laser.

**Figure 4**: Absorption spectroscopy of the $M_{11}$ optical transition on individual metallic SWNTs. Absorption spectra of individual metallic nanotubes assigned to species with (a): $2n + m = 27$ (gray: raw data, red: smoothed data); and (b): $2n + m = 24$ (gray: raw data, blue: smoothed data). (**c**): Histogram of the $M_{11}$ peak energies measured from 23 individual metallic nanotubes. The two groups corresponding to $2n + m = 27$ (red bars) and $2n + m = 24$ (blue bars) are distinguished.

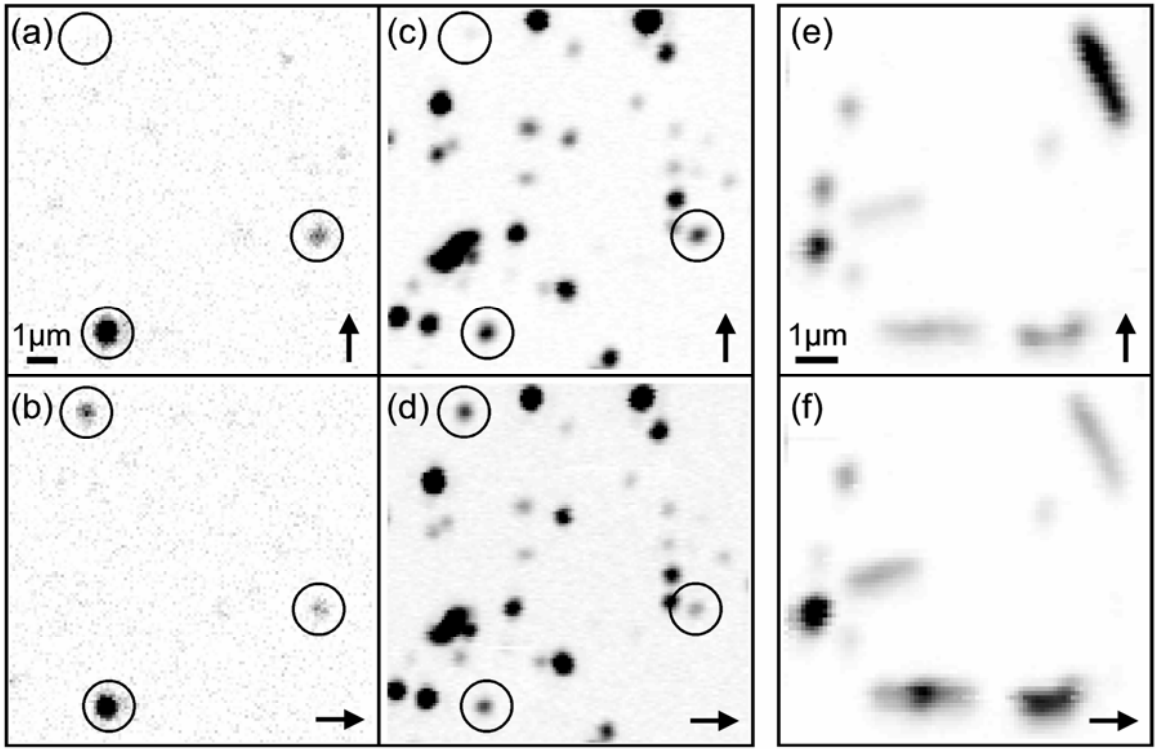

**Figure 1**

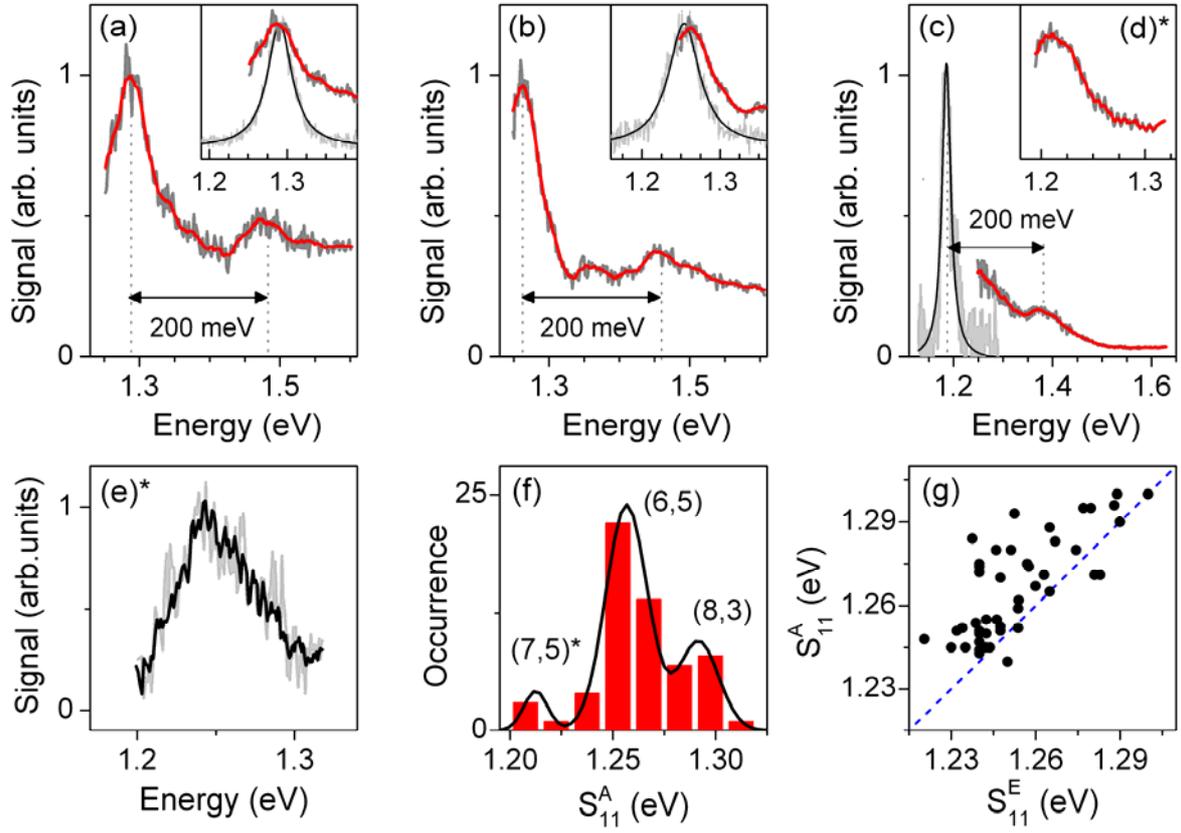

Figure 2

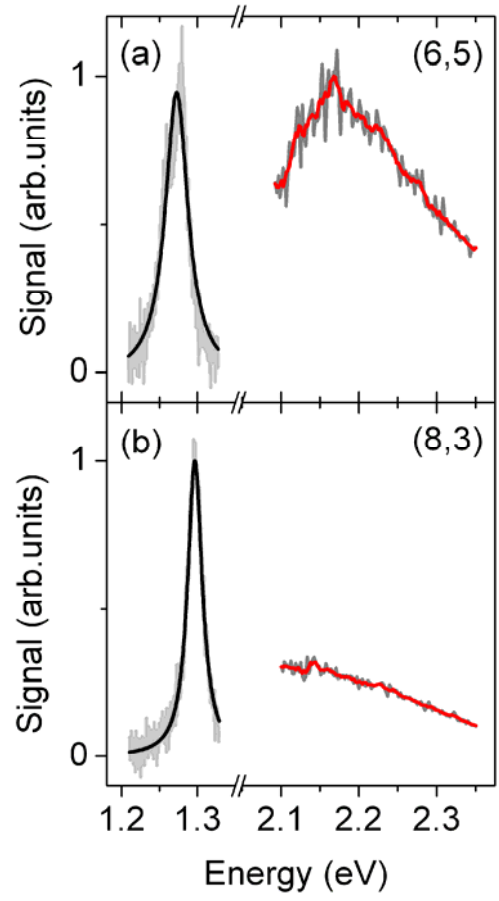

**Figure 3**

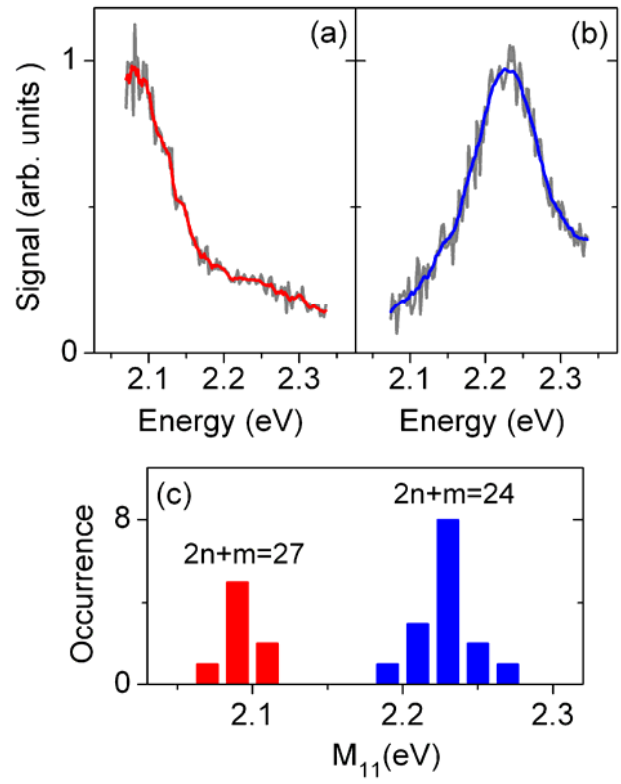

**Figure 4**